\documentclass[10pt,letterpaper]{article}
\usepackage{opex3}

\usepackage{graphicx}
\usepackage{epstopdf}

\usepackage{cite}

\usepackage{amsmath}

\usepackage{amsfonts}

\usepackage[colorlinks,urlcolor=blue,linkcolor=blue,citecolor=blue,anchorcolor=blue]{hyperref}

\begin{document}

\title{Controlling multi-wave mixing signals via photonic band gap of electromagnetically induced absorption grating in atomic media}

\author{Yiqi Zhang, Zhenkun Wu, Xin Yao, Zhaoyang Zhang, Haixia Chen, Huaibin Zhang, Yanpeng Zhang$^*$}

\address{Key Laboratory for Physical Electronics and Devices of the Ministry of Education \& Shaanxi Key Lab of Information Photonic Technique, Xi¡¯an Jiaotong University, Xi¡¯an 710049, China China}

\email{$^*$Corresponding author: ypzhang@mail.xjtu.edu.cn} 

\homepage{http://ypzhang.gr.xjtu.edu.cn} 


\begin{abstract}
We experimentally demonstrate dressed multi-wave mixing (MWM) and
the reflection of the probe beam due to electromagnetically induced absorption (EIA) grating can coexist in a five-level atomic ensemble.
The reflection is derived from the photonic band gap (PBG) of EIA grating,
which is much broader than the PBG of EIT grating.
Therefore, EIA-type PBG can reflect more energy from probe than EIT-type PBG does,
which can effectively affect the MWM signal.
The EIA-type as well as EIT-type PBG can be controlled by multiple parameters including the frequency detunings,
propagation angles and powers of the involved light fields.
Also, the EIA-type PBG by considering both the linear and third-order nonlinear refractive indices is also investigated.
The theoretical analysis agrees well with the experimental results.
This investigation has potential applications in all-optical communication and information processing.
\end{abstract}



\section{Introduction}

Electromagnetically induced transparency (EIT) can reduce the linear absorption of a beam when it passes through a resonant medium \cite{boller_prl_1991,gea-banacloche_pra_1995,hemmer_ol_1995},
and EIT also plays an important role in generating multi-wave mixing (MWM) processes \cite{kang_pra_2004,lukin_prl_1999}.
EIT enhanced four-wave mixing (FWM) in rubidium vapor \cite{hemmer_ol_1995,jain_prl_1996,zhang_prl_2007},
the efficient six-wave mixing (SWM) in a five-level close-cycled atomic system \cite{kang_pra_2004},
and coexisting of FWM and SWM in two ladder-type EIT windows \cite{zhang_prl_2007} have been investigated experimentally.
It is very interesting that EIT can change into electromagnetically induced absorption (EIA) \cite{artoni_prl_2006,brown_ol_2005,akulshin_pra_1998,lezama_pra_1999,akulshin_jop_2005,fuchs_jop_2007}
when the coupling field changes from traveling wave to standing wave,
which will induce electromagnetically induced grating (EIG) \cite{cardoso_pra_1999,cardoso_pra_2002,wen_apl_2011,zhangyiqi_ieee_2012} that can reflect the probe field \cite{brown_ol_2005,mitsunaga_pra_1999}.

The reflection of the probe field is a result of photonic band gap (PBG) structure of the EIG \cite{zhang_prl_2011,wan_pra_2011,zhang_lpl_2013}.
In previous literatures, the discussed PBG only appears at the EIT places \cite{andre_prl_2002,artoni_prl_2006,petrosyan_pra_2007,wu_josab_2008,cui_oe_2010,gao_ol_2010,wang_prl_2013},
and the width of the EIT-type PBG is less than 1 MHz.
As there is strong absorption in EIA region, so it is very hard to observe the EIA-type PBG.
However, the EIA region is so wide that if there is PBG, it will very be much wider than EIT-type PBG,
which would enhance the MWM signals through the Bragg reflection.
From this point of view, we plan to construct a configuration to observe the EIA-type PBG,
which is not been reported ever before.

In this paper, we experimentally and theoretically study the transmission as well as the reflection of the probe beam, and the MWM signals in the atomic ensemble.
In the emission direction of MWM signals, there coexist two signals:
the coherent emission (CE) signal in the MWM process and the electromagnetically induced Bragg reflection (EIBR) of the probe beam due to EIG.
As the PBG structure of the EIG can be affected by the frequency detunings, powers, and angles related with the pump beams,
we theoretically investigate the variation of the EIT-type and EIA-type PBGs by considering the above mentioned influence factors.
In experiment, we detect the probe as well as the MWM signals, analyze the intensity variations of the signals by using the PBG theory£¬
and demonstrate the existence of EIA-type PBG.

The organization of the paper is that we briefly introduced the basic theory in Sec. \ref{sec2},
display the experimental results and the explanations in detail in Sec. \ref{sec3},
and conclude the paper in Sec. \ref{sec4}.

\section{Basic Theory}\label{sec2}

\begin{figure}
  \centering
  \includegraphics[width=0.8\textwidth]{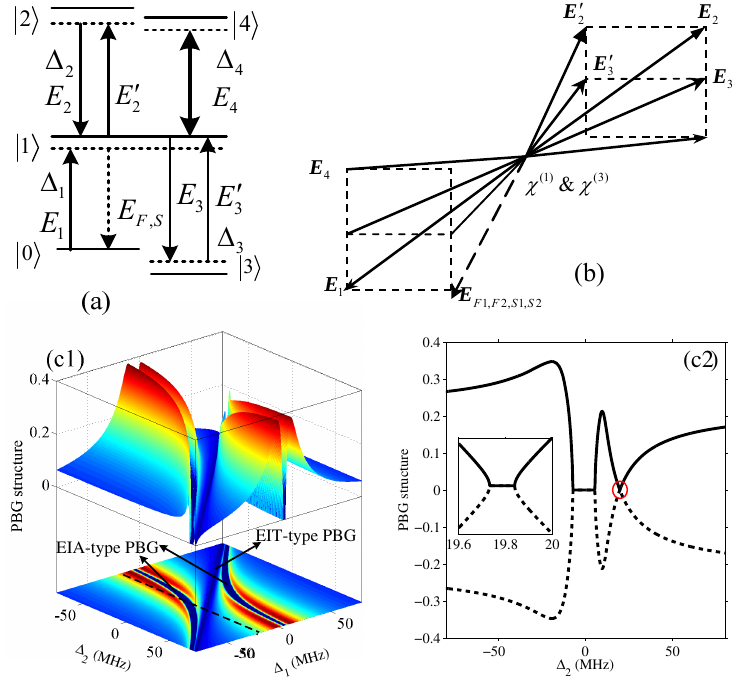}\\
  \caption{(a) Diagram of relevant energy levels.
  (b) Spatial beam geometry.
  (c) Theoretical PBG according to Eqs. (\ref{eq1})-(\ref{eq3}) when only linear susceptibility is considered.
  (c1) Theoretical three-dimensional PBG versus $\Delta_1$ and $\Delta_2$. The bottom panel is the top-view of the structure.
  (c2) PBG versus $\Delta_2$ corresponding to the dashed line in (c1), and the inset is the amplified view of the EIT-type PBG marked in the red circle.}
  \label{fig1}
\end{figure}

We consider an multi-level atomic system as shown in Fig. \ref{fig1}(a),
where the transition $|0\rangle \rightarrow |1\rangle$ is driven by one probe field $E_1$ (with frequency $\omega_1$, wavevector {\bf k}$_1$, Rabi frequency $G_1$, and power $P_1$),
the transition $|1\rangle \leftrightarrow |2\rangle$ is driven by two pump fields $E_2$ ($\omega_2$, {\bf k}$_2$, $G_2$, $P_2$)\&$E'_2$ ($\omega_2$, {\bf k}$'_2$, $G'_2$, $P'_2$),
the transition $|1\rangle \leftrightarrow |4\rangle$ is driven by $E_4$ ($\omega_4$, {\bf k}$_4$, $G_4$, $P_4$),
and the transition $|1\rangle \leftrightarrow |3\rangle$ is driven by $E_3$ ($\omega_3$, {\bf k}$_3$, $G_3$, $P_3$)\&$E'_3$ ($\omega_3$, {\bf k}$'_3$, $G'_3$, $P'_3$).
The Rabi frequency is defined as $G_i=\mu_{ij}E_i/\hbar$, in which $\mu_{ij}$ is the electric dipole moment between levels $|i\rangle$ and $|j\rangle$.
The spatial geometry of these beams is displayed in Fig. \ref{fig1}(b),
in which $E_2$\&$E'_2$ with a small angle between them ($\sim0.3^\circ$) propagate oppositely to $E_1$,
$E_3$\&$E'_3$ copropagate with but respectively deviate a small angle from $E_2$\&$E'_2$,
and $E_4$ copropagate with $E_2$ with a small angle between them.
With $E_3$\&$E'_3$ off, we can observe two FWM signals $E_{F1}$ ($\omega_1$, {\bf k}$_{F1}$) and $E_{F2}$ ($\omega_2$, {\bf k}$_{F2}$),
satisfying the phase matching conditions {\bf k}$_{F1}=${\bf k}$_1$+{\bf k}$_2-${\bf k}$'_2$ and {\bf k}$_{F2}=${\bf k}$_1$+{\bf k}$_4-${\bf k}$_4$, respectively.
In addition, when $E_3$\&$E'_3$ are open, there will be two SWM signals $E_{S1}$ and $E_{S2}$ as well as another FWM signal $E_{F3}$,
which satisfy the phase-matching conditions {\bf k}$_{S1}=${\bf k}$_1$+{\bf k}$_3-${\bf k}$'_3$+{\bf k}$_4-${\bf k}$_4$,
{\bf k}$_{S2}=${\bf k}$_1$+{\bf k}$_3-${\bf k}$'_3$+{\bf k}$_2-${\bf k}$'_2$,
and {\bf k}$_{F3}=${\bf k}$_1$+{\bf k}$_3-${\bf k}$'_3$, respectively.
In experiment, the signals are detected by photomultiplier tubes that are displayed along the output directions of the signals.

The two couple of pump beams $E_2$\&$E'_2$ and $E_3$\&$E'_3$ can interfere with each other to form two standing waves, i.e.,
$|{G_2}{|^2} = G_2^2 + {G'_2}{}^2 + 2{G_2}{G'_2}\cos (2{k_2}x)$ and $|{G_3}|^2 = G_3^2 + {G'_3}{}^2 + 2{G_3}{G'_3}\cos (2{k_3}x)$
to offer periodical dressing effects. The periodic standing wave also induces EIG,
which will lead to EIBR of $E_1$ because of the inertial PBG structure.
So, the detected MWM signal includes EIBR and CE.
According to the dressed perturbation chain method \cite{sun_pra_2004},
the total susceptibility of the system can be written as
\begin{equation}\label{eq1}
\begin{split}
  \chi  = &i\frac{{N\mu _{10}^2}}{{\hbar {\varepsilon _0}}}\left( {\frac{1}{{{d_1} + |{G_2}{|^2}/{d_2}}} - \frac{1}{{d'_1 + |{G_3}{|^2}/{d_3}}}} \right) \\
  &- i\frac{{N\mu _{10}^2}}{{\hbar {\varepsilon _0}}}\left( {\frac{{G_2^2 + {G'_2}^2}}{{{d_2}{{({d_1} + |{G_2}{|^2}/{d_2})}^2}}} - \frac{{G_3^2 + {G'_3}^2}}{{{d_3}{{(d'_1 + |{G_3}{|^2}/{d_3})}^2}}}} \right),
\end{split}
\end{equation}
in which $d_1=\Gamma_{10}+i\Delta_1$,
$d_2=\Gamma_{20}+i(\Delta_1+\Delta_2)$,
$d'_1=\Gamma_{1}-i\Delta_1$,
$d_3=\Gamma_{03}-i(\Delta_1+\Delta_3)$,
$\varepsilon_0$ is the permittivity in free space,
$\Gamma_{ij}$ is the population decaying rate between levels $|i\rangle$ and $|j\rangle$,
$\Delta_i=\Omega_{ij}-\omega_i$ is the detuning,
and $\Omega_{ij}$ is the transition frequency between $|i\rangle$ and $|j\rangle$.
In Eq. (\ref{eq1}), the terms in the two brackets are the linear ($\chi^{(1)}$,
$\rho _{00}^{(0)} \xrightarrow{\omega_1} \rho _{10}^{(1)}$ and $\rho _{00}^{(0)} \xrightarrow{-\omega^*_1} \rho _{01}^{(1)}$)
and third-order nonlinear ($\chi^{(3)}|E_{2,3}|^2$,
$\rho _{00}^{(0)} \xrightarrow{\omega_1} \rho _{10}^{(1)} \xrightarrow{\omega_2} \rho _{20}^{(2)} \xrightarrow{-\omega_2} \rho _{10}^{(3)}$ and
$\rho _{00}^{(0)} \xrightarrow{-\omega^*_1} \rho _{01}^{(1)} \xrightarrow{\omega_3} \rho _{31}^{(2)} \xrightarrow{-\omega^*_3} \rho _{01}^{(3)}$)
susceptibilities, respectively, where $\rho_{ij}$ is the matrix density element.
Based on the multi-parameter controllability of the system, $\Delta_2 \approx \Delta_3$ and $k_2 \approx k_3$ can be obtained to some extent.
Thus, the imaginary part of $\chi$ in Eq. (\ref{eq1}) will be significantly reduced.
Here, we note that the third-order nonlinear susceptibilities in $\chi$ cannot be considered if the pump field intensities are not high
(e.g., cw laser beams instead of pulsed laser beams are applied into the system).

As to the PBG of the EIG, we can obtain by adoptting the plane-wave expansion method \cite{wu_josab_2008,cui_oe_2010,wang_prl_2013}.
Expending $\chi$ as Fourier series and considering two-mode approximation,
we obtain the expression of the Bragg wavevector as follows
\begin{equation}\label{eq2}
  q =  \pm \frac{1}{2k}\sqrt{\left[ k^2(1 + c_0) - k^2 \right]^2 - k_1^4c_1^2},
\end{equation}
where $c_0$ and $c_1$ are the first two Fourier coefficients of $\chi$ Fourier series;
$k$ represents $k_2$ or $k_3$.
In the following text, we will never distinguish $k_2$ from $k_3$, and $\Delta_2$ from $\Delta_3$,
because of their similar values. This approximation is feasible for calculation.
The expressions of $c_0$ and $c_1$ can be written as

\begin{subequations}\label{eq3}
\begin{equation}\label{eq3a}
  {c_0} = i\frac{{N\mu _{10}^2}}{{{\varepsilon _0}\hbar }}\left[ {\left( {\frac{A}{{\sqrt {1 - {B^2}} }} - \frac{{A'}}{{\sqrt {1 - {{B'}{}^2}} }}} \right) - \left( {\frac{C}{{\sqrt {{{(1 - {B^2})}{}^3}} }} - \frac{{C'}}{{\sqrt {{{(1 - {{B'}{}^2})}{}^3}} }}} \right)} \right],
\end{equation}
\begin{equation}\label{eq3b}
  {c_1} = i\frac{{N\mu _{10}^2}}{{{\varepsilon _0}\hbar }}\left[ {\left( {A\frac{{\sqrt {1 - {B^2}}  - 1}}{{B\sqrt {1 - {B^2}} }} - A'\frac{{\sqrt {1 - {{B'}{}^2}}  - 1}}{{B'\sqrt {1 - {{B'}{}^2}} }}} \right) + \left( {\frac{{BC}}{{\sqrt {{{(1 - {B^2})}{}^3}} }} - \frac{{B'C'}}{{\sqrt {{{(1 - {{B'}{}^2})}{}^3}} }}} \right)} \right],
\end{equation}
\end{subequations}
where
\begin{align}
A =& \frac{d_2}{G_2^2 + G'_2{}^2 + {d_1}{d_2}},\quad B = \frac{2{G_2}{G'_2}}{G_2^2 + G'_2{}^2 + {d_1}{d_2}},\notag\\
C =& \frac{(G_2^2 + G'_2{}^2){d_2}}{{(G_2^2 + G'_2{}^2 + {d_1}{d_2})^2}},\quad A' = \frac{d_3}{G_3^2 + G'_3{}^2 + {d'_1}{d_3}},\notag\\
B' =& \frac{2{G_3}{G'_3}}{G_3^2 + G'_3{}^2 + {d'_1}{d_3}},\quad C' = \frac{(G_3^2 + G'_3{}^2){d_3}}{(G_3^2 + G'_3{}^2 + {d'_1}{d_3})^2}.\notag
\end{align}
Similar to the distribution of terms in Eq. (\ref{eq1}),
the terms in the two brackets in Eqs. (\ref{eq3a}) and (\ref{eq3b}) correspond to the linear and third-order susceptibilities, respectively.
According to Eqs. (\ref{eq2}) and (\ref{eq3}), the PBG structure in three-dimension when only the linear susceptibility is considered is shown in Fig. \ref{fig1}(c1),
which is scaled in color -- blue represents the smallest value while red the biggest value.
The bottom panel in Fig. \ref{fig1}(c1) shows the top-view of the three-dimensional PBG structure, from which the blue regions (i.e. the PBG) are more readable.
Corresponding to the dashed line shown in Fig. \ref{fig1}(c1) the PBG versus $\Delta_2$ is displayed in Fig. \ref{fig1}(c2),
the inset in which indicates the amplified EIT-type PBG. From Fig. \ref{fig1}(c),
it is clear to see that the EIA-type PBG is almost 110 times wider than EIT-type PBG.
The PBG width connects tightly with the term under the square root in Eq. (\ref{eq2}).
The necessary but not sufficient condition is $[k^2(1+c_0) k^2]^2-k_1^4c_1^2 \leq 0$.
Thus, we end up with the following formula,
\begin{equation}\label{eq4}
  {\Delta _{{\rm{gap}}}} = \frac{{({\Delta _1} + {\Omega _{10}})[2 + 2{{\rm Re}} ({c_0}) + {{\rm Re}} ({c_1})]}}{{2\sqrt {1 + {{\rm Re}} ({c_0})} \left( {1 + \frac{1}{2}{{\left[ {\frac{{{{\rm Im}} ({c_0})}}{{1 + {{\rm Re}} ({c_0})}}} \right]}^2}} \right)}}.
\end{equation}
If we introduce the substitutions $\omega_0=\Delta_1+\Omega_{10}$ and ${n_0} = \sqrt {1 + {\rm Re} ({c_0})} $,
Eq. (\ref{eq4}) can be rewritten as
\begin{equation}\label{eq5}
  {\Delta _{{\rm{gap}}}} = \frac{{{\omega _0}n_0^3[2n_0^2 + {\rm Re} ({c_1})]}}{{2n_0^4 + {{\rm Im}^2}({c_0})}},
\end{equation}
by using which we can analyze the changing of EIT-type as well as EIA-type PBG width.
Just take the case in Fig. \ref{fig1}(c) for example.
As known, imaginary part of the susceptibility connects with the gain/absorption.
At the EIT place, the nonlinearity contrast is large, and meanwhile the absorption is suppressed,
so a narrow-band PBG can be achieved.
Commonly, in the EIA region, the absorption is extremely high which cannot induce the appearance of PBG,
even though the nonlinearity contrast is much larger than that of EIT.
In our scheme, the imaginary part of the susceptibility is much reduced, and corresponding real part is doubled,
so the ${\rm Im}(c_0)$ in Eq. (\ref{eq5}) becomes very small that the PBG corresponding to EIA appears with a much large width according to Eq. (\ref{eq5}).
Considering that the EIA-type PBG is much wider than EIT-type PBG,
much more probe will be transformed into FWM in the EIA region than at the EIT place based on the EIBR mechanism.

Also, the absorption of the probe photon will generate fluorescence (FS) signal.
In this system, without any filter, the detected FS signal is composed of three components:
the single-photon FS in the decay $|1\rangle \rightarrow |0\rangle$ (R1),
two-photon FS in the decays $|2\rangle \rightarrow |1\rangle$ (R2) and $|4\rangle \rightarrow |2\rangle$ (R3).
So, the energy conservation for the system is
\begin{equation}\label{eq6}
  \frac{1}{I_{in}}(I_{R1}+I_{R2}+I_{R3}+I_M+R+T)=1,
\end{equation}
in which the terms are the intensities of FS, MWM, BR/transmission of the probe and input probe intensities, respectively.

\section{Experimental Results and Analysis}\label{sec3}
\subsection{Frequency detuning modulated PBG}

We first execute our experiment in the system composed by the states $5S_{1/2}(F=3)$ $(|0\rangle)$, $5P_{3/2}~(|1\rangle)$,
$5D_{5/2}$ $(|2\rangle)$, $5D_{3/2}$ $(|4\rangle)$ and $5S_{1/2}(F=2)$ $(|3\rangle)$ in $^{85}$Rb \cite{steck_alkali},
and the temperature of the atomic cell is about 220$^\circ \rm C$, which corresponds to an atomic density of $10^{12}$ cm$^{-3}$.
From four external cavity diode lasers (ECDL), the wavelengths of $E_1$, $E_2$\&$E'_2$, $E_3$\&$E'_3$ and $E_4$ are 780 nm, 775.98 nm, 795 nm and 776.16 nm, respectively.
The widths of laser beams are all about 100 $\mu$m.
Because ECDL is a cw laser that the high-order nonlinear susceptibilities will not be considered because of not high enough beam powers.

\begin{figure}
  \centering
  \includegraphics[width=\textwidth]{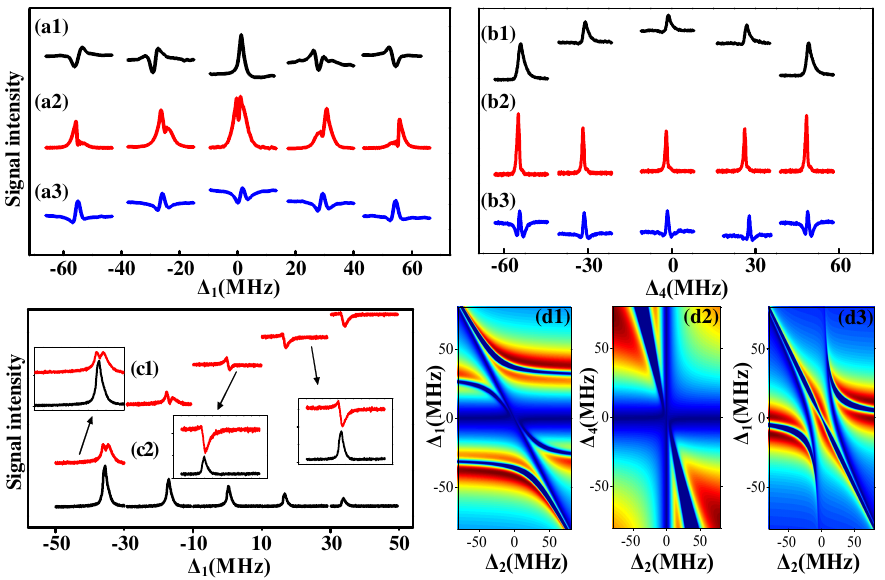}\\
  \caption{(a1)-(a3) Measured PT, FWM, and FS signal versus $\Delta_2$ for different $\Delta_1$, respectively.
  (b1)-(b3) Similar to (a1)-(a3) but versus $\Delta_2$ for different $\Delta_4$.
  (c1) and (c2) Measured PT and FWM signal with a larger power of $E_1$ and dressing effects from $E_2$ and $E_4$.
  (d1)-(d3) Theoretical PBGs versus $\Delta_1$\&$\Delta_2$ and $\Delta_2$\&$\Delta_4$ correspond to (a)-(c), respectively.}
  \label{fig2}
\end{figure}

In Fig. \ref{fig2}(a), we show the experimental results versus $\Delta_2$ with different $\Delta_1$.
With $\Delta_2$ changing from $\Delta_1+\Delta_2=0$, near resonant, to far away from resonant along the positive direction successively,
the probe transmission (PT) shows a single peak, decreasing left-peak and increasing right-dip, and a single dip correspondingly (Fig. \ref{fig2}(a1)).
Because a single EIT, reducing left-EIT and rising right-EIA, and single EIA occur at the above places successively.
In addition, the concomitant PBGs are tiny, left-narrow and right-wide, and large along the positive $\Delta_1$ direction, respectively,
which would reflect and decrease the probe to make the dip in the PT more obvious.
The case for $\Delta_1$ changing along the negative direction can be understood similarly as above.

For the FWM signal $E_{F1}$, it shows two enhancement peaks and one suppression dip when $\Delta_1$ is scanned as shown by the background profile in Fig. \ref{fig2}(a2),
which can be also theoretically demonstrated by $\rho _{10}^{(3)}$.
When $\Delta_2$ is scanned with $\Delta_1$ being fixed, $E_{F1}$ will theoretically show enhancement peak,
half-enhancement peak and half-suppression dip, suppression dip, half-suppression dip and half-enhancement peak,
and last an enhancement peak when $\Delta_1$ is changing from a large negative value to a large positive value.
However, all the curves exhibited in Fig. \ref{fig2}(a2) show two peaks, which seems contradict to the theoretical prediction mentioned above.
The reason is because of the BR of PBG as shown in Fig. \ref{fig2}(d1) that is according to Eq. (\ref{eq1}) from the plane-wave expansion method.
And the wider the PBG, the more BR from probe will be added into $E_{F1}$,
thus the BR can fill the suppression dip to lead to a small peak.
For the case $\Delta_1=0$, the dip is too deep for BR to fill up that there is still a small dip left over in the profile.

It is worth mentioning that EIA-type PBG is much wider than EIT-type PBG as shown in Fig. \ref{fig2}(d1).
Because the refractive index contrast is much larger than the absorption,
but at the dark state the former is relatively small though the latter is significantly suppressed.
Considering that FS signal together with PT and FWM satisfies Eq. (\ref{eq2}),
the FS shown in Fig. \ref{fig2}(a3) is easy to understand.

Doubly dressing effect should be considered if we open $E_4$, and $K$ for $\chi^{(1)}$ will be modified as $K=d_1+|G_2|^2/d_2+|G_4|^2/d_4$ with $d_4=\Gamma_{40}+i(\Delta_1+\Delta_4$).
Thus, the Fourier coefficients in Eqs. (\ref{eq3a}) and (\ref{eq3b}) should be replaced by the doubly dressed ones.
In Fig. \ref{fig2}(b), PT, FWM and FS signals versus $\Delta_2$ with different $\Delta_4$ and $\Delta_1=0$ are exhibited.
The background of PT shows an EIT peak because of the satisfaction of the two photon resonance (TPR) condition $\Delta_1+\Delta_4=0$,
and the PTs versus $\Delta_2$ also show EIT peaks are also because the TPR condition $\Delta_1+\Delta_2=0$ is satisfied.
Just from the doubly dressed $\chi^{(1)}$ in which $\Delta_2$ and $\Delta_4$ play a similar role,
one can also predict that the background PT versus $\Delta_4$ as well as the PTs versus $\Delta_2$ should show similar profiles.
The height of the PT versus $\Delta_2$ is the smallest because $K$ is the biggest (both the two TPR conditions are satisfied around $\Delta_4=0$).
The other condition is that EIBR from the PBG will also reduce the PT intensity.
As shown in Fig. \ref{fig2}(d2), the PBG becomes wider with bigger $\Delta_4$ that would transfer more energy from PT into FWM to fill up the dips and induce peaks as shown in Fig. \ref{fig2}(a2).
Because the PBG is the smallest around $\Delta_4=0$, the EIBR is also not big, thus the corresponding FWM is also the smallest.

If $E_1$ is large sufficiently to give dressing effect,
we study the tri-dressing effects as shown in Fig. \ref{fig2}(c).
Without the dressing effect from $E_1$, the PT will show a EIT peak because of the satisfaction of $\Delta_1+\Delta_2=0$ in $\chi^{(1)}$ around $\Delta_1=0$.
However, with $\Delta_1$ increasing from 0, the PT first shows a dip in a peak, then a peak and a dip, and last only a dip.
In order to understand the phenomenon, we introduce $|G_1|^2/[\Gamma_{00}+|G_2|^2/(\Gamma_{21}+i\Delta_2)]$ into $K$ besides the other two dressing fields.
By modifying the Fourier coefficients $c_0$ and $c_1$ with the tri-dressing ones, we display the corresponding PBG in Fig. \ref{fig2}(d3).
In comparison with the EIT-type PBG in Fig. \ref{fig2}(d1), the PBG here is EIA-type because the dressing effect from $E_1$ divides EIT into two parts and shifts them.
Thus, there are two EITs and one EIA if $\Delta_2$ is scanned around $\Delta_1=0$.
Considering that the EIA-type PBG is much broader than the EIT-type,
the EIBR will transfer part of the PT intensity into FWM more here than that in Fig. \ref{fig2}(a).
So, a peak together with a dip will appear in the PT around $\Delta_1=0$.
If $\Delta_1$ is large enough, only EIA can be satisfied which will swallow the peak and dig a dip in the PT.

For the FWM, it will show a peak because of the enhancement condition around $\Delta_1=0$.
When $\Delta_1$ increases, the suppression and enhancement can coexist, that the peak decreases.
When $\Delta_1$ increases over a critical value, only enhancement condition can be satisfied,
and considering the enhancement is much weaker than that around $\Delta_1=0$,
the peak here is the smallest. We note here that the FWM in Fig. \ref{fig2}(c2) is always enhanced by the EIBR.
But, the PBG width (Eq. (\ref{eq5})) decreases with $\Delta_1$ increasing as shown in Fig. \ref{fig2}(d3),
that the EIBR from PT to FWM also reduces.
Even though there is a region in which EIT-type and EIA-type PBGs coexist that can lead to a quite large EIBR,
the suppression in this region still reduces the peak.

\subsection{Angle and power modulated PBG }

\begin{figure}
  \centering
  \includegraphics[width=1\textwidth]{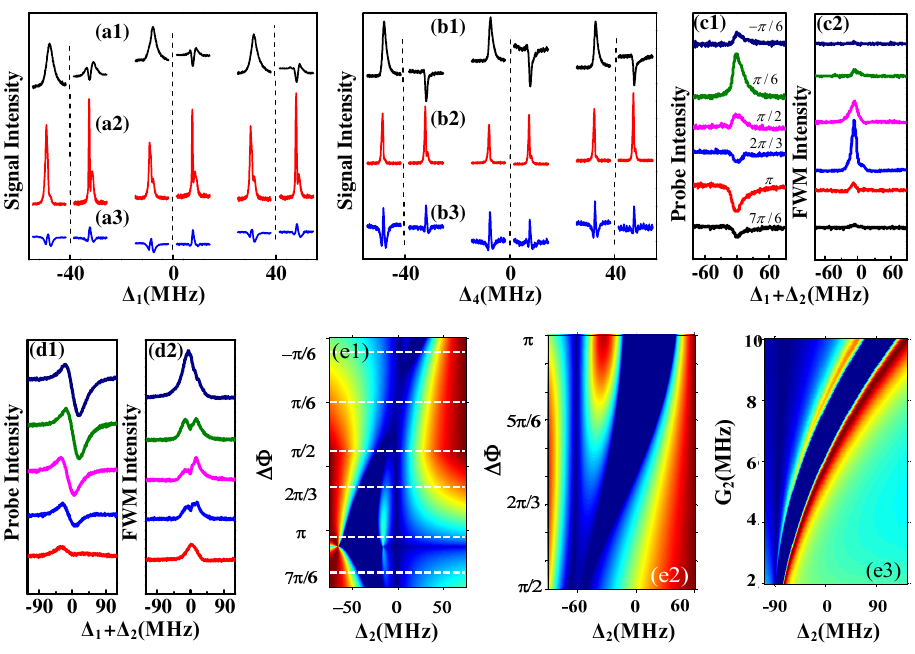}\\
  \caption{(a1)-(a3) PT, FWM and FS signals versus $\Delta_2$ with different $\Delta_1$ and $\Delta_4=0$.
  (b1)-(b3) Similar to (a1)-(a3) but with different $\Delta_4$ and $\Delta_1=0$.
  The other parameters are $P_1=$3.6 mW, $P_2=$30.6 mW, $P'_2=$5.4 mW and $P_3=$14.0 mW.
  (c1)-(c2) PT and FWM signals versus $\Delta_2$ with $\Delta_1$ fixed.
  (d1)-(d2) Similar to (c1)-(c2) but with $\Delta_1=$100 MHz when $E_1$, $E_2$ and $E'_2$ are turned on.
  Other parameters are $P_1=$4 mW and $P'_2=$8 mW with $P_2=$2.4, 8, 12, 16 and 20 mW from bottom to top.
  (e1)-(e3) Theoretical PBGs correspond to (a)-(d), respectively.}
  \label{fig3}
\end{figure}

If we introduce an additional phase factor $e^{i\Delta \Phi}$ to the dressing fields,
$K$ in $\chi^{(1)}$ for PT is modified as $K_1=d_1+|G_2|^2e^{i\Delta \Phi}/d_2$ (for single dressed) or $K_2=d_1+|G_2|^2e^{i\Delta \Phi}/d_2+|G_4|^2/d_4$ (for doubly dressed).
Here, $\Delta \Phi$ connects with the angle between the two incident beams, and it could be manipulated by the orientations of induced dipole moments \cite{brown_ol_2005}.
Correspondingly, the doubly dressed $E_{F1}$ which is proportional to $(1/K_2)^2$ and FS which is proportional to $1/[d_2+|G_2|^2e^{i\Delta \Phi}/(d_1+|G_4|^2/d_4)]$ can be also controlled by $\Delta \Phi$.
In addition, by adjusting $\Delta \Phi$, bright and dark states could switch into each other,
so that the PBGs associated with them would also be affected.
In this part, we only discuss the region around $\Delta_1=0$, in which the PT shows pure EIT peak. In each row of Fig. \ref{fig3}(a),
there are three couples of signals that are recorded with $\Delta \Phi = \pi/2$ (left) and $\Delta \Phi = 5\pi/6$ (right), respectively.
The configuration of Fig. \ref{fig3}(b) as well as that of Fig. \ref{fig3}(c) is same as that of Fig. \ref{fig3}(a).

In Fig. \ref{fig3}(a1), the PTs with $\Delta \Phi = \pi/2$ show pure peaks.
When $\Delta \Phi$ is changed into $5\pi/6$,
the PBG as shown in Fig. \ref{fig3}(e1) becomes wider and leads to more EIBR and dig a deep dip in the peak of the PT profile.
The corresponding FWM shown in Fig. \ref{fig3}(a2) at $\Delta \Phi = \pi/2$ shows a pure peak with slightly AT splitting because $K_1$ plays a slightly bigger role than $d_2$ in $\rho _{10}^{(3)}$.
When $\Delta \Phi = 5\pi/6$, the increased EIBR from PT significantly enhances the left peak of the AT splitting FWM.

Because the reflecting properties of the EIG depend on $\Delta \Phi$,
we can investigate the different angle-property for doubly dressed case with $\Delta_2$ scanned by changing $\Delta_4$, as shown in Fig. \ref{fig3}(b).
As analyzed in Fig. \ref{fig2}(b), the PTs with $\Delta \Phi = \pi/2$ shown in Fig. \ref{fig3}(b1) will show EIT peaks.
When $\Delta \Phi = 5\pi/6$, the width of the PBG increases sharply as shown in Fig. \ref{fig3}(e2) which will lead to the PT switch from a peak to a dip because of the quite large EIBR.
Considering that the EIBR will transfer into FWM, the FWM displayed in Fig. \ref{fig3}(b2) is enhanced when $\Delta \Phi$ is changed from $\pi/2$ to $5\pi/6$.
The discussion above can also explain the FS as shown in Fig. \ref{fig3}(b3),
the intensity of which can be considered as Iin subtracting the intensities of PT and FWM.

In order to investigate the angle-dependence,
we change $\Delta \Phi$ in a much wide range, from $-\pi/6$ to $7\pi/6$.
The experimental results are shown in Figs. \ref{fig3}(c1) and \ref{fig3}(c2),
which are the PT and FWM, respectively.
We can see that the PT first shows a small peak,
next to a high peak, then a small peak, then a small dip, then a deep dip,
and last a small dip from top to bottom.
Taking the corresponding EIA-type PBG (the EIT-type case can be neglected in comparison with the EIA-type one) shown in Fig. \ref{fig3}(e1)
into account, as indicated by the white dashed lines, the changing of PBG width is in a small, smaller, small, high, higher, and high manner from top to bottom.
Because of the idea that the wider the PBG the more EIBR, the PT changing in Fig. \ref{fig3}(c1) is reasonable.
As to the corresponding FWM, as shown in Fig. \ref{fig3}(c2),
can be also explained in the way discussed above, that the wide PBG will enhance the FWM due to the EIBR which can be modulated by $\Delta \Phi$.
However, in the Fig. \ref{fig3}(c2), although the PBG width at $\Delta \Phi = \pi$ or $7\pi/6$ is obvious broader than that at $\Delta \Phi = 2\pi/3$ (see the Fig. \ref{fig3}(e1)),
the FWM signal is much smaller, which contradicts the analysis debated above.
The reason is that $\Delta_2 \approx \Delta_3$ can be fulfilled theoretically,
however it cannot exactly obtained in experiment when we scanning $\Delta_2$.
Thus, the EIBR will be reduced greatly if $\Delta_2 \approx \Delta_3$ cannot be satisfied very well.

The powers of dressing and probe fields also can affect the PBG.
In Figs. \ref{fig3}(d1)-\ref{fig3}(d3), with $\Delta_1$ fixed and $P_2$ being very low,
a single-peak structure with a not obvious dip in the PT versus $\Delta_2$ can be observed.
With $P_2$ increases, the peak moves leftward and a dip appears on right of the peak.
Further increasing $P_2$, the dip becomes wider and deeper, and the peak also becomes higher gradually.
In the transmission probe as shown in Fig. \ref{fig3}(d1),
the peak (dip) higher (lower) than baseline is EIT (EIA) induced by $E_2 (E'_2)$ through the term $|G_2|^2/d_2$ when $\Delta_2$ is scanned.
The corresponding bright state is at $\Delta_1+\Delta_2|G_2|^2/|G_1|^2=0$ and dark state at $\Delta_2 = 0$.
With the power of $E_2 (E'_2)$ increased, the height of EIT (EIA) gets higher (lower) as shown in Fig. \ref{fig3}(d1),
which leads to the enhancement of the FWM as shown in Fig. \ref{fig3}(d2) because of the increasing EIT-type as well as EIA-type PBG width which is shown in Fig. \ref{fig3}(e3).
At last, the EIA-type PBG is much wider than the EIT-type one, that the two peaks of FWM merge into one wide peak (the top curve in Fig. \ref{fig3}(c2)).

\subsection{Modulated EIA-type PBG in SWM}

\begin{figure}
  \centering
  \includegraphics[width=\textwidth]{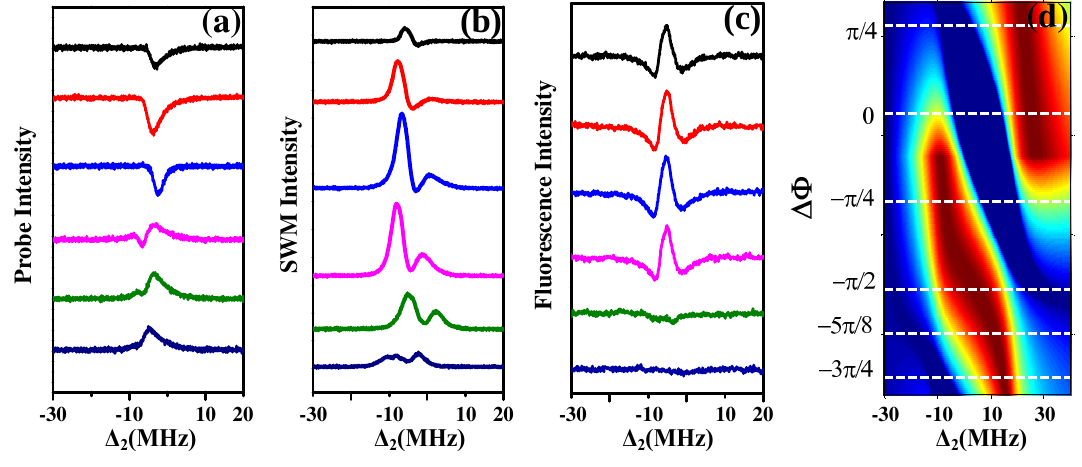}\\
  \caption{PT (a), SWM (b) and FS signals (c) versus $\Delta_2$ at $\Delta_1=-40$ MHz with
   $\Delta \Phi=-3\pi/4$, $-5\pi/8$, $-\pi/2$, $-\pi/4$, 0 and $\pi/4$ from the bottom to top, respectively.
  (d) Theoretical PBG versus $\Delta_2$ and $\Delta \Phi$, and the dashed lines correspond to the curves in (a).
  Other parameters are $P_1=$7 mW, $P_2=$17 mW, and  $P'_2=$8 mW.}
  \label{fig4}
\end{figure}

To further understand the EIA-type PBG,
we investigate the SWM signal $E_{S1}$ in the system by turning all fields.
Because of the multi-dressing effects,
the $K$ in $\chi^{(1)}$ will be modified into $d_1+e^{i\Delta \Phi}[|G_1|^2/(\Gamma_{00}+|G_2|^2/d_{21})+|G_2|^2/d_2+|G_4|^2/(d_{41}+|G_2|^2/d_{124})]$
with $d_{21}=\Gamma_{21}+i\Delta_2$, $d_{41}=\Gamma_{41}+i(\Delta_1+\Delta_4)$ and $d_{124}=\Gamma_{24}+i(\Delta_1+\Delta_2+\Delta_4)$.
We show the intensities of the probe transmission,
SWM and FS signals versus $\Delta_2$ (from negative to positive) with different $\Delta \Phi$  in Figs. \ref{fig4}(a)-\ref{fig4}(c), respectively.
The corresponding theoretical PBG is shown in Fig. \ref{fig4}(d).

Let's explore the PBG shown in Fig. \ref{fig4}(d) first.
Indicated by the dashed lines from bottom to top, the PBG width first decreases, and then increases;
when it reaches a maximum it decreases again.
Recalling the EIBR effect of PBG, it can be understood that the PT first shows a dip,
then a peak-dip structure, and last a peak along the increasing of $\Delta \Phi$ from bottom to top in Fig. \ref{fig4}(a).
Specifically, at $\Delta \Phi = -3\pi/4$, EIA-type PBG is narrow, so the EIBR is weak and the PT shows a peak.
With $\Delta \Phi$ changing from $-3\pi/4$ to $-\pi/2$, the width of EIA-type PBG decreases,
which leads the peak in PT to get larger. The peak-dip structure is for the reason that the PBG gets modulated as $\Delta \Phi$ is altered.
When $\Delta \Phi$ changes from $-\pi/2$ to $\pi/4$, the PBG is much broadened although PBG is narrow at $\Delta \Phi = \pi/4$,
which will lead to much stronger EIBR and the switch from peak into dip. As to the $E_{S1}$ signal shown in Fig. \ref{fig4}(b),
it can be understood by the changing of EIBR analyzed above.

For the FS signal shown in Fig. \ref{fig4}(c),
it can be determined by the PT and $E_{S1}$ signals according to Eq. (\ref{eq6}),
which is also affected by the changing of PBG.
For example, with the increasing of PBG, the PT decreases, while the sum of $E_{S1}$ and FS signals increases.

Next, we consider a five-level atomic system as shown in Fig. \ref{fig1}(a) is composed of energy levels
$3S_{1/2}(F=1)$ $(|0\rangle)$, $3P_{3/2}$ $(|1\rangle)$,
$4D_{3/2}$ $(|2\rangle)$, $5S_{1/2}$ $(|4\rangle)$ and $3S_{1/2}(F=2)$ $(|3\rangle)$ of sodium \cite{steck_alkali}.
The transitions $|0\rangle \rightarrow |1\rangle$,
$|1\rangle \rightarrow |2\rangle$,
$|1\rangle \rightarrow |4\rangle$,
$|1\rangle \rightarrow |3\rangle$
are driven by fields with wavelengths 589.0 nm, 568.8 nm, 616.1 nm, and 589.0 nm, respectively \cite{zuo_prl_2006}.
The atomic density as well as the width of the beams is similar to rubidium vapour as mentioned before.
Due to the pulsed laser beams are applied into the system, the third-order nonlinear susceptibility need to be considered.

\begin{figure}
  \centering
  \includegraphics[width=0.8\textwidth]{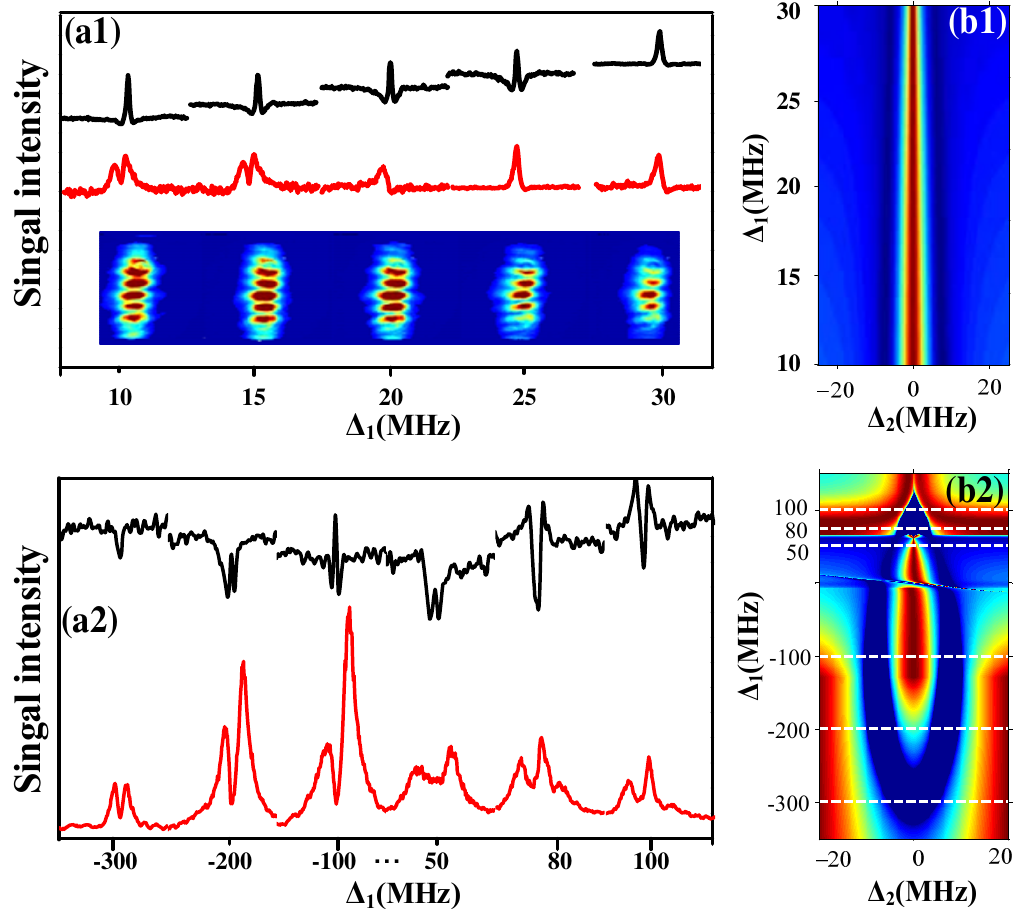}\\
  \caption{(a1) PT (first row), FWM (second row) signals versus $\Delta_2$ and PT images (third row) with $\Delta_1=$10, 15, 20, 25, and 30 MHz from left to right, respectively.
  (a2) Similar to (a1) but with $\Delta_1$ far away from TPR. The $x$--labels just display the values of $\Delta_1$, at which the experimental results are obtained (have noting to do with the $x$--scale).
  (b1) and (b2) Theoretical PBGs correspond to (a1) and (a2), respectively.}
  \label{fig5}
\end{figure}

In Figs. \ref{fig5}(a1) and \ref{fig5}(a2), the PTs (first row) as well as FWM (second row) versus $\Delta_2$ with different $\Delta_1$ around and away from the resonance region are shown, respectively.
The corresponding theoretical simulations of PBG are shown in Figs. \ref{fig5}(b1) and \ref{fig5}(b2), respectively.

When $\Delta_1$ is set $\Delta_1=10$, 15 and 20 MHz (away from $\Delta_1=0$),
a dip and a very tiny peak within this dip can be seen in the PT as shown in Fig. \ref{fig5}(a1).
If only the linear susceptibility $\chi^{(1)}$ is considered, only an EIA can be obtained,
but if the third-order nonlinear susceptibility $\chi^{(3)}$ is taken into consideration,
an EIT with narrower width appears in the center of EIA (the three left-up curves in Fig. \ref{fig5}(a1)).
Here, EIA is concomitant with wide PBG as shown in the Fig. \ref{fig5}(b1),
and the effect of the PBG concomitant with EIT can be neglected.
As discussed above, the wider PBG due to EIA can result in strong FWM signal via both enhancing the CE and EIBR components.
However, the EIT will lead to a dip in this FWM peak via suppressing both the two components,
as shown by the three left-low curves in Fig. \ref{fig5}(a1).
However, as shown by the two right-up curves in Fig. \ref{fig5}(a1), with $\Delta_1$ increasing,
the wide dip in the PT becomes shallower ($\Delta_1=25$ MHz) and even disappears ($\Delta_1=30$ MHz),
but the peak within it becomes higher. The reason is that EIA-type PBG width decreases with $\Delta_1$ increasing as depicted by Fig. \ref{fig5}(b1),
which leads to the decreasing of EIBR of the probe and enhancement of the PT.
The disappearance of the wide dip at $\Delta_1=30$ MHz is because that the EIT due to $\chi^{(1)}$ becomes stronger,
while the EIA due to the combined effect of $\chi^{(1)}$ and $\chi^{(3)}$ becomes weaker.
Therefore, there is only single peak in the PT.
However, though the EIBR component decreases,
the CE component is not influenced because the atomic coherence effect is still significant in this region,
so the peaks of FWM is not completely corresponding to the dips in the PT signal as described by the two right-low curves in Fig. \ref{fig5}(a1).

PT and FWM affected by the PBG can be also verified by the splitting of the probe images,
which connects with the nonlinear phase shift by $\phi_{NL}\propto1/I_1$ \cite{zhang_pra_2009}.
There are two reasons responsible for such phenomena.
First, according to the relative position between the probe $E_1$ and the strong dressing $E_2$ ($E'_2$), $E_1$ overlaps with $E_2$ and $E'_2$ in $y$ direction,
which leads to the splitting of $E_1$ in $y$ direction due to the attraction of $E_2$ and $E'_2$ beam. Second,
when $\Delta_1$ changes from 10 MHz to 30 MHz, the intensity of PT is affected by the modulated PBG due to changing of $\Delta_1$,
i.e, the PT increases with the increasing of $\Delta_1$ due to the decreasing PBG width.
Therefore, with $I_1$ increasing, the splitting of $E_1$ image weakens, as shown by the images in Fig. \ref{fig5}(a1).

Figure \ref{fig5}(a2) shows the PT (top row) and FWM signal (bottom row) versus $\Delta_2$  but with $\Delta_1$ far away from TPR when $\chi^{(1)}$ and $\chi^{(3)}$ are considered simultaneously.
The corresponding PBGs are shown by the dashed lines in Fig. \ref{fig5}(b2),
in which the EIT is indicated by the slope region inbetween the dashed lines at $\Delta_1=-100$ MHz and $\Delta_1=50$ MHz.
As discussed above, the PT shown in Fig. \ref{fig5}(a2) can be easily understood according to the PBG sketch in Fig. \ref{fig5}(b2).
For example, the dip in the PT at $\Delta_1=-300$ MHz can be explained by the wide EIA-type PBG indicated by the bottom dashed line in Fig. \ref{fig5}(b2).
Corresponding to the PT, FWM shown in Fig. \ref{fig5}(a2) can be also understood through the PBG width together with the discussions mentioned above.

\section{Conclusion}\label{sec4}
In summary, we experimentally and theoretically investigated the enhancement of the MWM signals by the EIBR from EIT- and EIA-type PBGs in atomic vapors by considering two EIGs simultaneously. We found that the EIA-type PBG is much wider than the EIT-type one, so the former one can bring more EIBR and make the MWM signal at EIA place be enhanced significantly. Such enhancement effect can be affected by the frequency detunings, powers, and the related angles of the involved fields. The theory and experimental results agree with each other very well. The investigation provides a novel perspective on PBG formed in atomic vapors, and has potential applications in all-optical communications as well as signal processing.

\section*{Acknowledgement}
This work was supported by
the 973 Program (2012CB921804), CPSF (2012M521773),
NSFC (61308015, 61078002, 61078020, 11104214, 61108017, 11104216, 61205112),
RFDP (20110201110006, 20110201120005, 20100201120031),
and FRFCU (xjj2013089, 2012jdhz05, 2011jdhz07, xjj2011083, xjj2011084, xjj2012080).
\end{document}